\documentclass[preprint,prl,amsmath,amssymb]{revtex4-1}

\usepackage{graphicx}% Include figure files
\usepackage{mathptmx}

\usepackage{color}
\usepackage{ulem}

\begin{document}

\title{Understanding the origin of band gap formation in graphene on metals: graphene on Cu/Ir(111)}

\author{H. Vita,$^1$ S. B\"ottcher,$^1$ K. Horn,$^1$ E. N. Voloshina,$^2$ R. E. Ovcharenko,$^2$ Th. Kampen,$^3$ A. Thissen,$^3$ and Yu. S. Dedkov$^{3,}$\footnote{Corresponding author. E-mail: Yuriy.Dedkov@specs.com}}
\affiliation{$^1$Fritz-Haber-Institut der Max-Planck-Gesellschaft, 14195 Berlin, Germany}
\affiliation{$^2$Institut f\"ur Chemie, Humboldt Universit\"at zu Berlin, Brook-Taylor-Str. 2, 12489 Berlin, Germany}
\affiliation{$^3$SPECS Surface Nano Analysis GmbH, Voltastra\ss e 5, 13355 Berlin, Germany}

\date{\today}

\begin{abstract} 
\textbf{Understanding the nature of the interaction at the graphene/metal interfaces is the basis for graphene-based electron- and spin-transport devices. Here we investigate the hybridization between graphene- and metal-derived electronic states by studying the changes induced through intercalation of a pseudomorphic monolayer of Cu in between graphene and Ir(111), using scanning tunnelling microscopy and photoelectron spectroscopy in combination with density functional theory calculations. We observe the modifications in the band structure by the intercalation process and its concomitant changes in the charge distribution at the interface. Through a state-selective analysis of band hybridization, we are able to determine their contributions to the valence band of graphene giving rise to the gap opening. Our methodology reveals the mechanisms that are responsible for the modification of the electronic structure of graphene at the Dirac point, and permits to predict the electronic structure of other graphene-metal interfaces.}
\end{abstract}

\maketitle

The discovery of the unique transport properties of graphene~\cite{Novoselov:2005,Zhang:2005} has stimulated the search for practical applications of this material in devices based on the transport of electrical charge or/and spin. In most of these devices, the graphene-metal interfaces are utilized as a contact for charge or spin injection. Hence the crystallographic and electronic properties of such interface determine the charge/spin injection efficiency of the corresponding contact, which is one of the main motivations for studies of the graphene-metal interface~\cite{Wintterlin:2009,Batzill:2012,Dedkov:2012book}. 

From another, more fundamental point of view, the study of the bonding mechanism at the graphene-metal interface is a very interesting problem in itself, and up to now such systems are far from the being fully understood~\cite{Voloshina:2012c}. Several factors influence the electronic properties of the graphene-metal interface: charge transfer from/onto graphene-derived $\pi$ states, hybridization of the electronic valence band states of graphene and the metal, and the lattice match between the graphene and metal surface. These factors determine the behaviour of the $\pi$ states in the vicinity of the Fermi level, $E_F$, (as, e.\,g., a deviation from the linear behaviour characteristic for free-standing graphene) as well as the appearance of an energy gap in the spectrum of the graphene-derived electronic states at the so-called Dirac point, $E_D$. From the point of view of electronic structure, two scenarios may be distinguished~\cite{Voloshina:2012c}: those in which the characteristic linear dispersion near $E_D$ is largely preserved, such as on Ir(111), Pt(111), and Cu(111), and those where a massive rearrangement of bands occurs, such as Ni(111), Co(0001), Ru(0001), for example.

Intercalation of metals in between graphene and substrates offers an interesting scientific playground to investigate the metal-graphene interaction. First, graphene may become decoupled from strongly interacting substrates, such as in the case of Au or Al intercalation in between graphene and nickel~\cite{Shikin:2000,Dedkov:2001,Varykhalov:2008,Voloshina:2011NJP}. Such intercalated layers may also change the carrier concentration in graphene, and even change the carrier type (from electrons to holes) such as in the case of Au intercalation in graphene/SiC(0001)~\cite{Gierz:2010}. Intercalated metals may also enhance the magnetic coupling between a ferromagnetic substrate and graphene, with a view to utilizing graphene as a spin filter~\cite{Weser:2010,Weser:2011}. Moreover, the passivating, protecting function of graphene may also be used in such systems~\cite{Dedkov:2008d,Dedkov:2008e}. The intercalated layer in itself may bring new properties to graphene, such as in the case of lithium where superconductivity has been predicted to occur~\cite{Profeta:2012hg}.
  
The different factors that determine the bonding between graphene and metal can be analyzed by studying graphene-based intercalation systems, and recently, such attempts were undertaken for graphene on Pt(111) and Ir(111), and on pseudomorphic layers of Co and Ni grown on these substrates~\cite{Gyamfi:2012eg,Decker:2013ch,Pacile:2013jc}. On top of Pt(111) or Ir(111), graphene has properties almost like the free-standing phase, as judged by its electronic structure derived from photoemission and scanning tunnelling spectroscopy~\cite{Sutter:2009a,Pletikosic:2009}. Intercalation of 1\,ML of Co or Ni leads to a strong buckling of the graphene layer similar to the one formed on Ru(0001)~\cite{Marchini:2007}, and a large energy gap between $\pi$ and $\pi^*$ states around the $K$ point occurs, due to the broken symmetry for the two carbon sublattices in the graphene unit cell, induced by the strong hybridization of the graphene $\pi$ and Co, Ni\,$3d$ valence band states; in both cases the linear dispersion of the graphene-derived states in the vicinity of $E_F$ is not conserved~\cite{Gyamfi:2012eg,Decker:2013ch,Pacile:2013jc}.

A very different situation is found for intercalation of noble metals (Cu, Ag, Au) and the formation of single close-packed pseudomorphic layers. Here, due to the absence of $d$ states in the close vicinity of $E_F$ and the change of the doping of graphene upon intercalation, the influence of $\pi-d$ hybridization effects on the electronic structure of graphene is much weaker ~\cite{Shikin:2000,Dedkov:2001,Dedkov:2003,Varykhalov:2010a}, and the linear dispersion of the graphene-derived $\pi$ states survives. The intercalation of single layers of noble metals is a suitable model system to investigate the interaction between graphene and metals, since it permits to follow the competition between the different (substrate and intercalated layer) electronic states.

Here we study a pseudomorphic single layer of Cu atoms intercalated in between graphene and Ir(111), using angle-resolved photoelectron spectroscopy (ARPES), scanning tunnelling microscopy (STM), and state-of-the-art density functional theory (DFT) calculations. The interaction between the metal and graphene can be studied in detail because of the sharp and well separated signatures of the Cu\,$3d$ bands in photoemission. Our results yield a complete picture of hybridization between the Cu and carbon derived states and the consequent opening of a band gap in the graphene $\pi$ bands; they are of importance for the understanding of the bonding mechanism at graphene-metal interfaces in general. This system is also interesting since it permits the creation of a single layer of Cu under considerable tensile strain, because of the lattice mismatch of ~6.2\% between the two metals. If deposited on top of an Ir (111) surface, this strain leads to de-wetting and the creation of 3D islands/clusters even at low coverages, or to dendritic growth of layers as in the case of Au on Ir(111)~\cite{Ogura:2007by} or incomplete growth of Cu on Pt(111)~\cite{Moras:2012hu}. 

\section*{Results}

\begin{figure}
\centering
\includegraphics[scale=1.2]{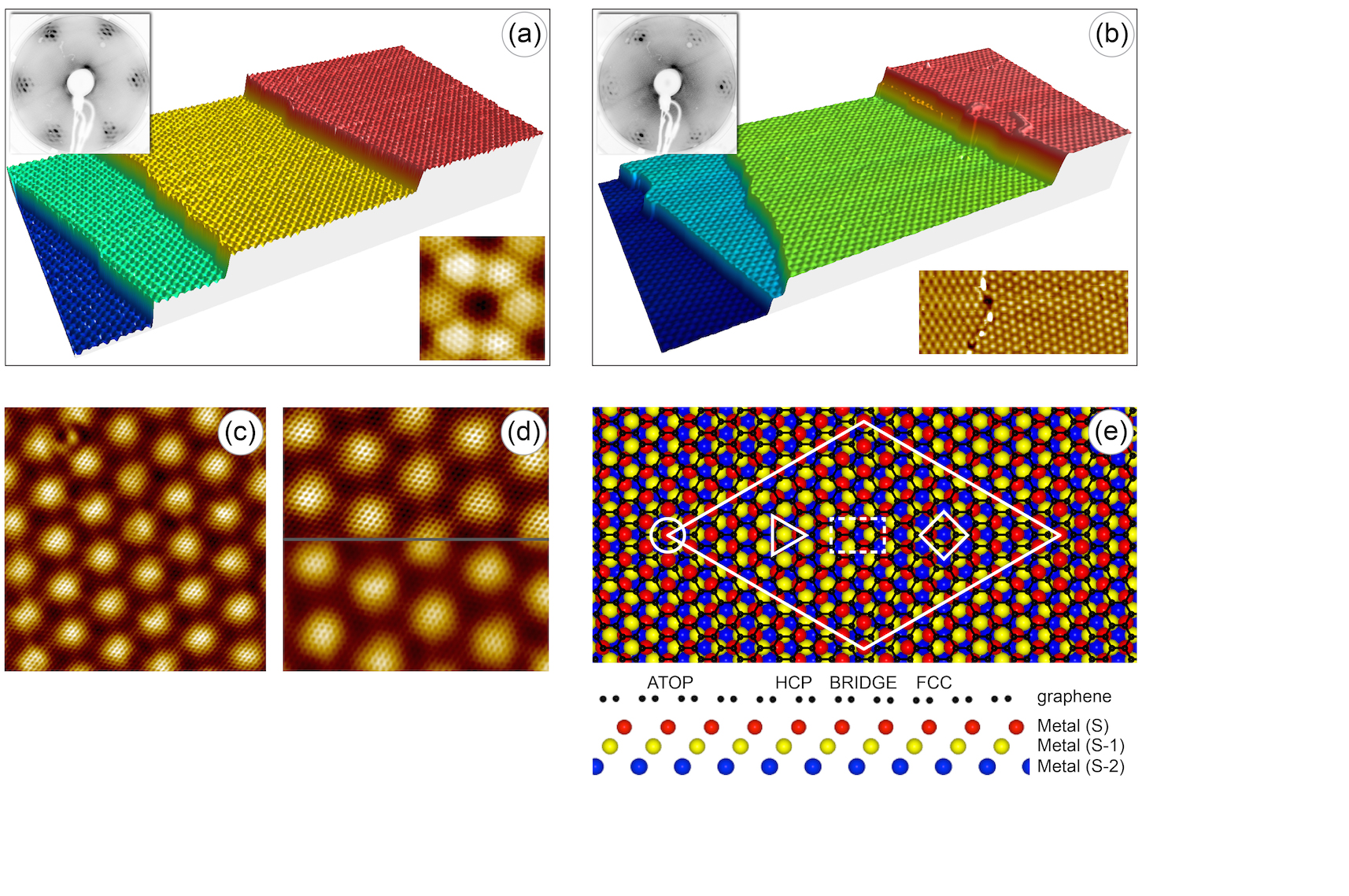}\\
\caption{\label{STM} \textbf{STM images of graphene on Ir(111) and pseudomorphic Cu/Ir(111).} (a) 3D view of the STM image of graphene/Ir(111) (scanning parameters: $120\times56$\,nm$^2$, $U_T=+0.3$\,V, $I_T=1.6$\,nA). Upper and lower insets show the LEED image ($E_p=93$\,eV) and atomically resolved STM image (scanning parameters: $4.5\times4.5$\,nm$^2$, $U_T=+0.5$\,V, $I_T=10$\,nA) of this system. (b) 3D view of the STM image of graphene/Cu/Ir(111) (scanning parameters: $175\times82$\,nm$^2$, $U_T=+0.3$\,V, $I_T=1.6$\,nA). Upper and lower insets show the LEED image ($E_p=90$\,eV) and an STM image of two graphene domains (scanning parameters: $60\times24$\,nm$^2$, $U_T=+0.3$\,V, $I_T=1.6$\,nA). (c) and (d) present atomically resolved images of the graphene/Cu/Ir(111) system obtained at $U_T=+0.3$\,V and $U_T=-0.3$\,V(top)/$-0.9$\,(bottom), respectively (scanning parameters: (c) $13.5\times13.5$\,nm$^2$, (d) $10.1\times10.1$\,nm$^2$, $I_T=1.6$\,nA for both images). (e) Top and side views of the graphene/metal(111) structure. In case of Ir(111) all layers consist of the same atoms, whereas for the Cu/Ir(111) pseudomorphic system the metal (S) is the copper layer. In the top view, the circle, triangle, rhombus, and rectangular denote the ATOP, HCP, FCC, and BRIDGE positions for the carbon atoms, respectively. The big white rhombus marks the unit cell of the graphene/metal(111) system.} 
\end{figure}

Graphene layers on Ir(111) were prepared in a standard way via cracking of propylene gas at $1100^\circ$\,C. STM images demonstrate the long range ordering of the moir\'e structure over several hundreds nm as shown in Fig.~\ref{STM}(a); this image is an extract of the originally acquired $300 \times 300$\,nm$^2$ STM data set. The lower inset shows an atomically resolved STM image of the moir\'e unit cell of graphene/Ir(111), which demonstrates the periodicity of 10 graphene unit cells over 9 unit cells of Ir(111), consistent with the corresponding low energy electron diffraction (LEED) pattern shown in the upper inset of Fig.~\ref{STM}(a)~\cite{NDiaye:2008qq,Hattab:2012fq,Hamalainen:2013jj}. At the bias voltages used in the experiment ($U_T=+0.3$\,V and $U_T=+0.5$\,V), graphene/Ir(111) is imaged in the so-called \textit{inverted} contrast~\cite{NDiaye:2008qq,Voloshina:2013dq}, where the crystallographically highest ATOP positions are imaged as dark areas and the lowest HCP and FCC places are imaged as bright ones, demonstrating the strong influence of the electronic structure on STM contrast (for the definitions of the high-symmetry positions of the graphene/metal moir\'e structure see Fig.\,S1 in the supplementary material).

Intercalation of a 1\,ML-thick Cu layer underneath graphene was performed via stepwise annealing of a deposited layer of Cu (with a nominal thickness of $1.5$\,ML) on graphene/Ir(111). During this procedure, the intensity of the ``live'' C\,$1s$ and Ir\,$4f$ photoelectron spectra was taken as the intercalation proceeded as a function of annealing temperature [see Fig.\,S2(a,b) in the supplementary material]. This method permits to follow the formation of an intercalated Cu layer underneath graphene on Ir(111), and the absence of any additional low binding energy (BE) components in the C\,$1s$ spectra after intercalation indicates that the Cu layer is completely intercalated.

The effective intercalation of a thin Cu layer, and the formation of the graphene/Cu/Ir(111) occurs at $550^\circ$\,C, is identified via strong modifications of the C\,$1s$ and Ir\,$4f$ emission lines [Fig.\,S2(a,b)]. Intercalation leads to a shift of the C\,$1s$ peak to higher binding energy; it can be fitted with two components at $284.69$\,eV and $285.01$\,eV [Fig.\,S2(c)] [the integral intensity of the C\,$1s$ line is restored to the value equivalent to the one for graphene/Ir(111)]. In case of the Ir\,$4f$ spectra, the energy splitting between bulk (b) and interface (i) components is reduced from $537$\,meV to $463$\,meV, and the intensity of the interface component is strongly suppressed compared to that of the graphene/Ir(111) system [Fig.\,S2(d)]. 

Although the STM data from the graphene with an intercalated Cu layer [Fig.~\ref{STM}(b-d)] may appear similar to those from graphene/Ir(111) at first glance, there are clear differences in detail which reveal the effect of intercalation. The large scale STM image shown in (b) demonstrates the formation of a moir\'e structure on top of Cu/Ir(111), which has the same periodicity of $(10\times10)$graphene/$(9\times9)$Cu/Ir(111). Hence we conclude that a pseudomorphic Cu layer on Ir(111) is formed after intercalation. However, contrary to the results for graphene on Ir(111), a variation of the tunnelling voltage during STM imaging of graphene/Cu/Ir(111) does not lead to an inversion of the imaging contrast -- graphene is here always imaged in \textit{direct} contrast [Fig.~\ref{STM}(c,d)]. The two atomically-resolved STM images were acquired at (c) $U_T=+0.3$\,V and (d) $U_T=-0.3$\,V (upper part) and $U_T=-0.9$\,V (lower part). In Fig.~\ref{STM}(d) the tunnelling voltage was changed during scanning ``on-the-fly'' showing the absence of any change of the contrast in the moir\'e cell on an atomic scale. 

Figure~\ref{STM}(e) shows the structure of the graphene/Cu/Ir(111) system as derived from DFT calculations. A $(10 \times 10)$ unit cell graphene lattice was placed on a $(9 \times 9)$ five layer Ir(111) oriented slab, and the graphene and the topmost Ir layers were then allowed to relax. The resulting in-plane lattice constants of graphene are close to those of the free-standing species. For the structural evaluation of the intercalated system, the topmost metal layer (S) in Fig.~\ref{STM}(e) was replaced by Cu. The Cu atoms were found to take up the positions of the Ir atoms, and the distance between the topmost metal layer and graphene in the ATOP position was found to be $3.581$\,\AA\ for graphene on Ir and $3.122$\,\AA\ for the intercalated Cu layer, a considerable reduction, signaling an enhanced interaction of carbon with the Cu atoms. Other structural parameters are given in Table\,T1 in the supplementary material. 

The conservation of long range periodicity of the system after intercalation of Cu is also shown by the LEED images of the graphene/Cu/Ir(111) system [upper inset of Fig.~\ref{STM}(b)]. However, while the periodicity of the spots reflecting the existence of the graphene moir\'e structure is similar to that for graphene/Ir(111), the symmetry of spot intensities is reduced from sixfold to threefold. Hence we directly infer a lowering of the local symmetry, which means that the sublattice symmetry for the two carbon atoms in the graphene unit cell is broken. Thus we expect a strong modification of the electronic structure of graphene in the vicinity of $E_F$.

\begin{figure}
\centering
\includegraphics[scale=1.2]{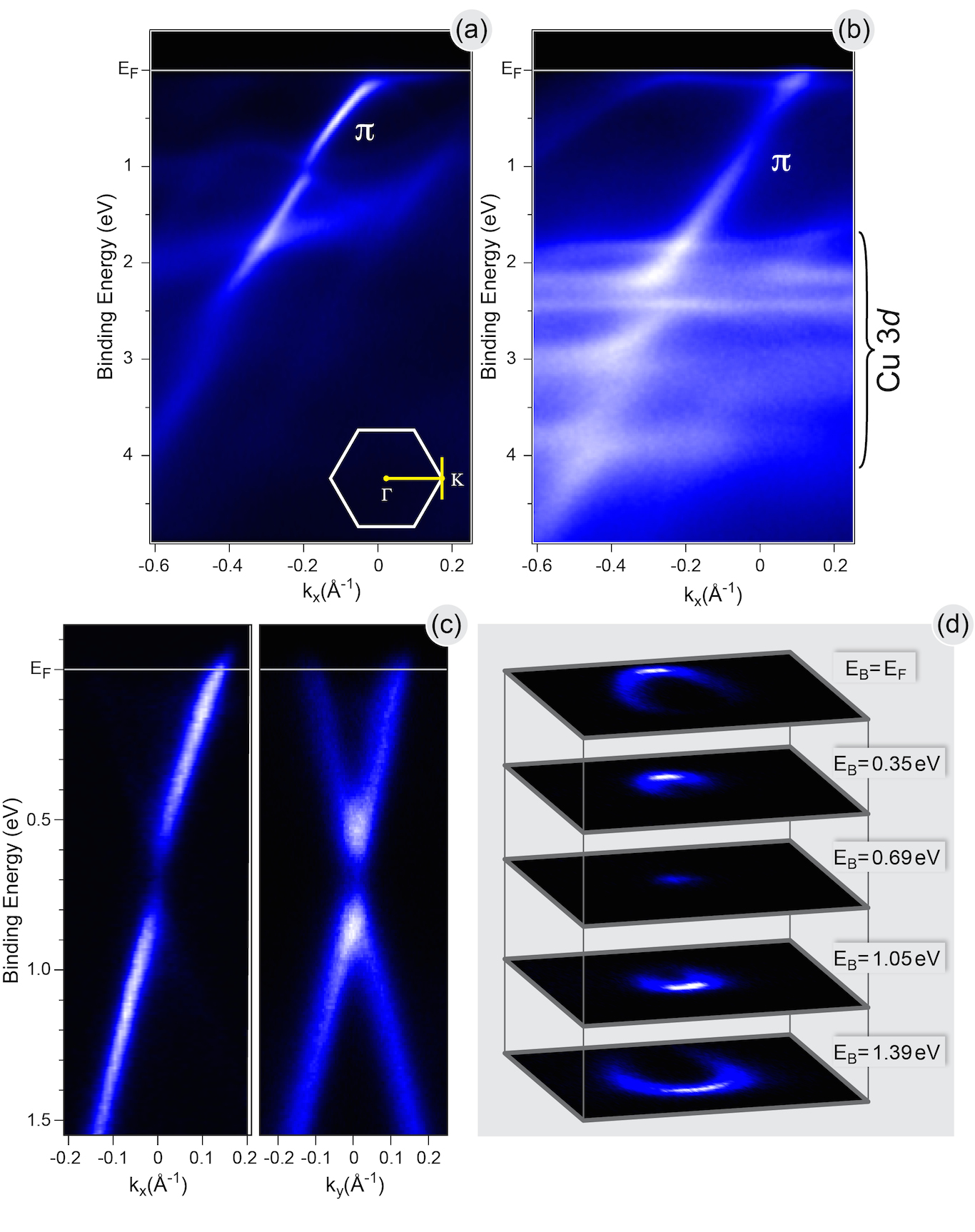}\\
\caption{\label{ARPES} \textbf{Electronic structure of graphene/Ir(111) and graphene/Cu/Ir(111) measured by ARPES.} (a,b) ARPES intensity maps for the graphene layer on Ir(111) and Cu/Ir(111), respectively, acquired along the $\Gamma-\mathrm{K}$ direction of the BZ of graphene with a photon energy $h\nu=65$\,eV. The inset in (a) shows a graphene-derived BZ with the corresponding directions. (c) ARPES intensity maps for graphene/Cu/Ir(111) obtained in the vicinity of the $\mathrm{K}$ point along $\Gamma-\mathrm{K}$ (left) and perpendicular to it (right) directions of BZ of graphene with photon energy $h\nu=40.81$\,eV. (d) The corresponding constant energy cuts extracted from the complete 3D ARPES data set for the graphene/Cu/Ir(111) system.} 
\end{figure}

This modification is revealed in detail using ARPES and a comparison of the results with our DFT calculations. Such data demonstrate deviations from the band structure expected for free-standing or electronically decoupled graphene, and because of the sharp and well-separated Cu $3d$ band spectral features, permit to identify details of band hybridization. Consider the photoemission intensity maps of graphene/Ir(111) and graphene/Cu/Ir(111) in the vicinity of $E_F$ in Fig.~\ref{ARPES}. For graphene/Ir(111) [Fig.~\ref{ARPES}(a)] the clear linear dispersion of the graphene $\pi$ states is observed around $E_F$ with energy gaps at higher binding energies due to avoided crossings among the main $\pi$ band and the replica bands which appear due to the additional periodicity of the moir\'e lattice~\cite{Pletikosic:2009}. By extrapolation we extract a position of the Dirac point at about $100\pm20$\,meV above $E_F$, corresponding to a slight $p$-doping of the graphene layer, in agreement with earlier ARPES data~\cite{Pletikosic:2009,Kralj:2011kq}. 

Intercalation of the Cu layer underneath graphene leads to a significant modification of the valence band states of the graphene layer [Fig.~\ref{ARPES}(b)]. Graphene on Cu/Ir(111) now is $n$-doped with a position of the Dirac point at $0.688$\,eV below $E_F$. Although the $\pi$ band still has a linear dispersion near $E_F$, a clear hybridization between the graphene $\pi$ and Cu\,$3d$ valence band states in the $2-4$\,eV binding energy region is obvious. This manifests itself as a series of avoided crossing gaps between graphene- and Cu-$d$-derived valence band states [see also Figure~\ref{HYBRID}]. Moreover, in the spectral function of the graphene $\pi$ band, an energy gap of $0.36$\,eV appears at the Dirac point. This energy gap is more clearly resolved in photoemission data sets obtained at a photon energy $h\nu=40.81$\,eV presented in Fig.~\ref{ARPES}(c,d), where ARPES intensity maps around the $\mathrm{K}$ point are shown in (c) for the $\Gamma-\mathrm{K}$ (left panel) and perpendicular to $\Gamma-\mathrm{K}$ (right panel) directions as well as (d) corresponding energy cuts at the respective binding energies marked in the figure. It is surprising that the present values for the energy of the Dirac point $E_D$ and the gap are quite different from those for graphene on bulk Cu(111) ($E_D-E_F= -0.3$\,eV, gap $=0.25$\,eV, \cite{Walter:2011fj}), and for a single intercalated Cu layer in between graphene and Ni(111) ($E_D-E_F = -0.3$\,eV, gap $=0.18$\,eV, \cite{Varykhalov:2010a}). The different width of the gap for the different graphene/Cu interface can be probably connected with the various periodicities of the corresponding moir\'e structures as was shown in Ref.~\cite{Song:2013ji}. Assuming the pseudomorphic growth of Cu layer at the interface in all cases the energy gap has to increase in the row graphene/Cu/Ni(111) $\rightarrow$ graphene/Cu(111) $\rightarrow$ graphene/Cu/Ir(111). 

\section*{Discussion}

The appearance of an energy gap around the $\mathrm{K}$ point in the spectrum of the graphene $\pi$ states is one of the intriguing problems in graphene interface studies, and its existence or absence, for example in graphene on SiC(0001), has been extensively discussed~\cite{Zhou:2008,Rotenberg:2008}. For single layer graphene on those metals that do not destroy the Dirac cone, i.\,e. the noble metals, Pt or Ir, gap openings have been observed in photoemission, and this has been attributed to sublattice symmetry breaking induced by a superstructure that couples the states at the $\mathrm{K}$ and $\mathrm{K'€™}$ points. As we will show through DFT calculations below, the dominating factor that determines the magnitude of the energy gap is the strength of hybridization of the graphene-derived $\pi$ states with the valence band states of the substrate, which leads to the broken symmetry of the two carbon sublattices as shown already by the LEED data in Fig.~\ref{STM}(b). 

\begin{figure}
\centering
\includegraphics[scale=1.2]{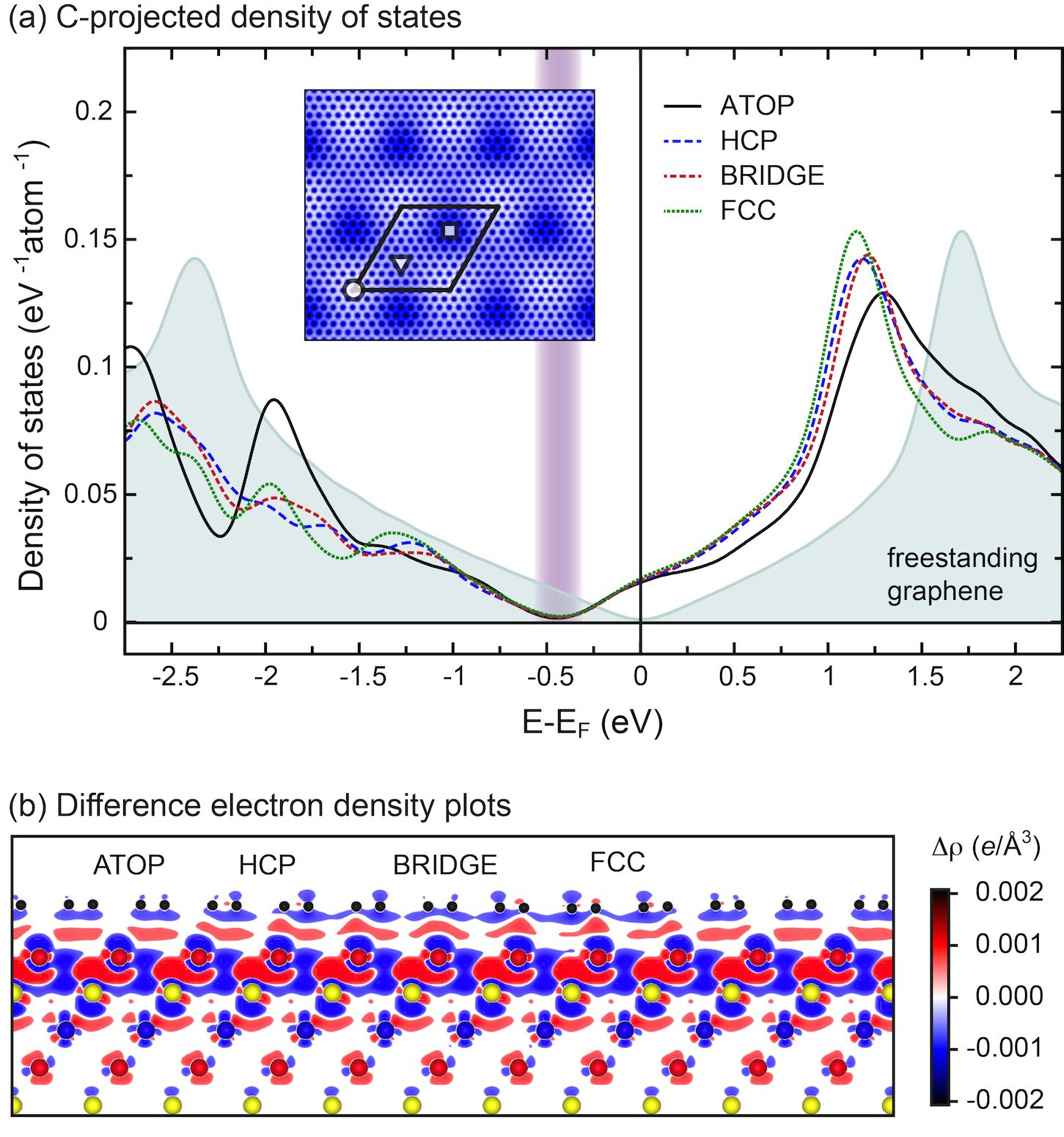}\\
\caption{\label{DOS_STM_CHARGE} \textbf{Results of the electronic structure DFT modelling for graphene/Cu/Ir(111).} (a) Carbon-atom projected site-resolved partial density of states for the graphene-derived $\pi$ states. Greyed plot shows the corresponding DOS for the free-standing graphene. The band gap region is marked by the shadow area. Inset shows the simulated STM image corresponding to the experimental results obtained at $U_T=-0.3$\,V. The high-symmetry positions are marked by the symbols similar to Fig.~\ref{STM}. (b) Side view of the graphene/Cu/Ir(111) system with the corresponding difference electron density, $\Delta\rho (r)=\rho_{gr/Cu/Ir(111)}(r)-\rho_{Ir(111)}(r)-\rho_{Cu(111)}(r)-\rho_{gr}(r)$, plotted in units of $\mathrm{e/\AA}^3$.} 
\end{figure}

The analysis of structural features is shown in Fig.~\ref{STM}(e). Here the interface metal layer (S) is either Ir or Cu on Ir(111). A pseudomorphic arrangement of the Cu layer on Ir(111) is assumed in the structure optimization procedure, a reasonable assumption in view of the LEED and STM data. At the energy minimum, a corrugation of the graphene layer on Cu/Ir(111) of $0.229$\,\AA\ is found, which is much lower than the one of $0.307$\,\AA\ for graphene/Ir(111)~\cite{Voloshina:2013dq}. The calculated carbon site projected partial density of states (PDOS) for the graphene-derived $\pi$ states in graphene/Cu/Ir(111) shows that graphene is $n$-doped with the Dirac point at $0.45$\,eV binding energy and a band gap of $0.15$\,eV [Fig.~\ref{DOS_STM_CHARGE}(a)], whereas the experiments give values of $0.688$\,eV and $0.36$\,eV, respectively [Fig.~\ref{ARPES}(c)]. The computed DOS was obtained by broadening of the originally calculated data for the density of states obtained for the graphene/Cu/Ir(111) supercell in the slab geometry: this broadening gives rise to a reduction of the energy gap, which would be 0.26\,eV from the unbroadened data (see Fig.\,S4 from supplementary material). A comparison of the calculated PDOSs for the Cu layer and the graphene layer indicates the existence of the hybridization between Cu\,$3d$ and graphene $\pi$ states in the binding energy range of $1.5-3.5$\,eV [Fig.~\ref{DOS_STM_CHARGE}(a) and Fig.\,S5 from supplementary material].

The effect of hybridization can be visualized in real space in the charge distribution across the moir\'e unit cell; the corresponding difference in electron density, $\Delta\rho (r)$, defined as a difference of densities for graphene/Cu/Ir(111) and those for the separate layers in the system, is shown in Fig.~\ref{DOS_STM_CHARGE}(b). The charge distribution picture for the FCC and HCP high symmetry sites is different from the one characteristic for graphene/Ir(111) and similar to the situation for graphene/Ni(111) or graphene/Rh(111) where a strong interaction for these positions is observed. However, here the effect is weaker because the hybridization occurs between the valence states at higher binding energies. This charge distribution was then used to model the STM images. A simulated image corresponding to a bias voltage of $U_T=-0.3$\,V is shown as an inset of Fig.~\ref{DOS_STM_CHARGE}, which is in good agreement with results presented in Fig.~\ref{STM}(d). As mentioned above, the ATOP positions of the graphene/Cu/Ir(111) system are imaged as bright areas corresponding to \textit{direct} imaging contrast; this situation prevails over a range from $+0.3$\,V and $-0.9$\,V [see Fig.\,S6 in the supplementary material]. This can be connected with the similar PDOS for different high-symmetry positions in the graphene moir\'e structure for this system [Fig.~\ref{DOS_STM_CHARGE}(a)] and that in this case the real topography contrast is prevailed in the STM images of graphene/Cu/Ir(111). 

\begin{figure}
\centering
\includegraphics[scale=1.2]{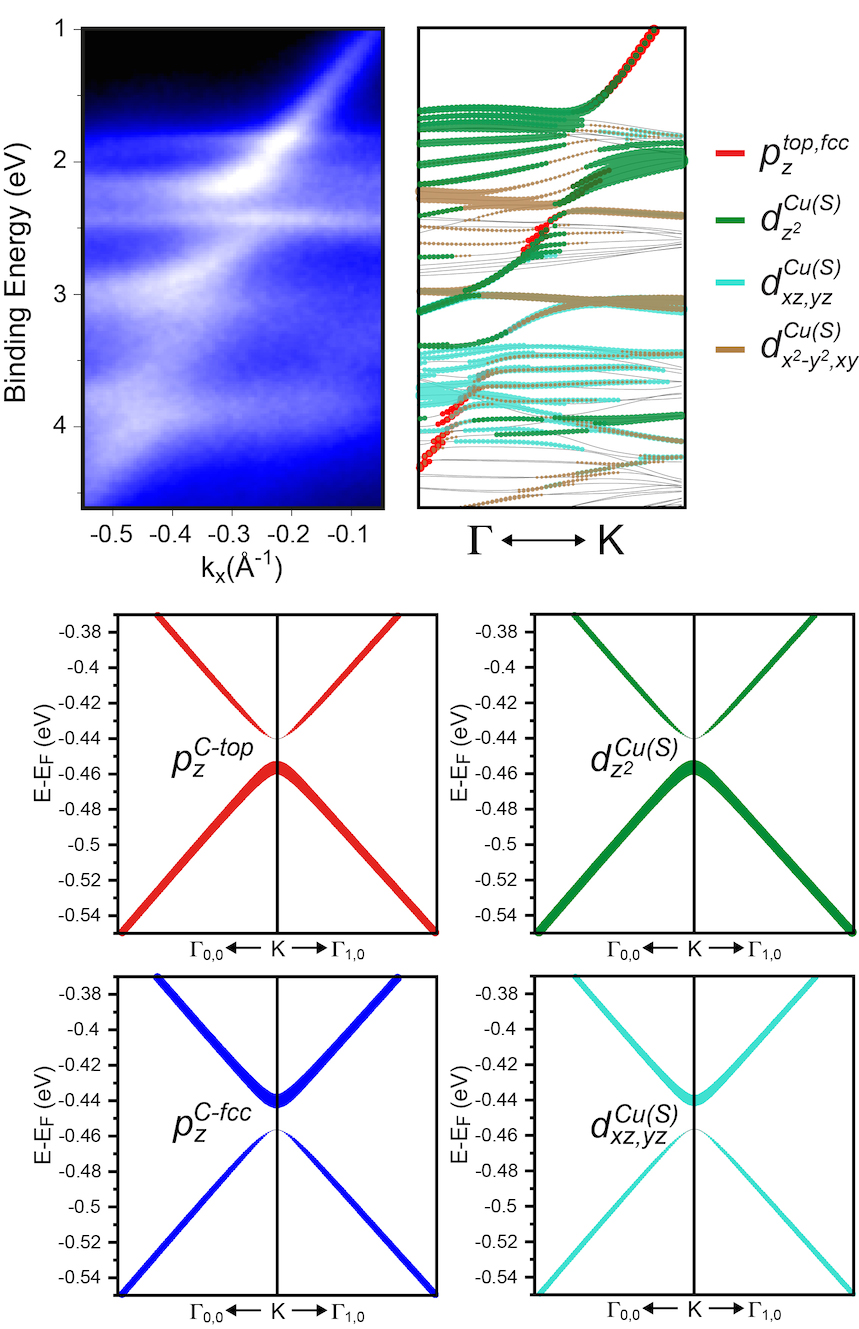}\\
\caption{\label{HYBRID} \textbf{Analysis of the electronic structure of graphene/Cu/Ir(111).} (Upper panel) Comparison of the the experimental and calculated electronic structure of graphene/Cu/Ir(111) and graphene/Cu(111), respectively, in the energy and $k$-vector ranges where hybridization of graphene $\pi$ and Cu $3d$ states is observed. (Lower panel) Orbital character decomposition of the valence band states of graphene in the vicinity of the Dirac point. See Fig.~\ref{STM}(e) and Fig.\,S1 in the supplementary material with the corresponding text for the definitions of the respective position of the carbon atoms.} 
\end{figure}

Because of the narrow line widths of the Cu $3d$-induced bands in photoemission, the graphene/Cu/Ir(111) system is an excellent one to study band hybridization. Five features altogether can be distinguished in the binding energy region from $1.8$\,eV to $4.0$\,eV. The avoided crossings of these bands with those derived from the graphene $\pi$ band are clearly observed [shown enlarged in Fig.~\ref{HYBRID}, upper panel, left], and the change of band character at around $2.15$\,eV, starting from mostly $d$-derived at $k = -0.6$\,\AA$^{-1}$ (with respect to the $\mathrm{K}$ point) to mostly graphene $p_z$-derived is evident (topmost grey feature). Moreover, there are clear differences in the amount of hybridization, reflected in the magnitude of the avoided crossings. Because of the computational cost of deriving the band structure of the complete system (large unit cell of the moir\'e structure), we restrict ourselves to an assignment of bands calculated for a Cu slab layer with the same lattice constant than graphene, i.\,e. $2.464$\,\AA. These data are shown next to the experimental bands in Fig.~\ref{HYBRID}. Because of the slab model, a multitude of bands appear in the calculations; the main weight of the bands are indicated by color bands. Their relation to the different atomic $d$ states from which they arise is indicated next to the calculation. The calculated bands are close to the experimental bands. 

Having shown the occurrence of hybridization of the graphene $\pi$ and Cu\,$3d$ valence band states through photoemission, we demonstrate that this process is responsible for the appearance of the energy gap in the electronic structure of graphene at the Dirac point. Hybridization leads to the intermixing of the $\pi$ orbitals with $3d$ orbitals of the corresponding orbital character (symmetry) depending on the carbon atom in the unit cell. According to the picture of Fig.~\ref{DOS_STM_CHARGE}(b), which shows the charge distribution difference, the most intensive interaction is observed around the FCC and HCP places. Here, one of the carbon atoms is placed above the interface Cu atom and hybridization between C$^{top}$\,$p_z$ and Cu\,$3d_{z^2}$ orbitals is observed. The second carbon atom is placed either above the $hcp$ or $fcc$ hollow site of Cu/Ir(111), and here the hybridization between C$^{fcc,hcp}$\,$p_z$ and Cu\,$3d_{xz,yz}$ orbitals occurs. For free-standing graphene the electronic states originating from two carbon sublattices are degenerate around the $\mathrm{K}$ point. The interaction of electronic states of carbon atoms with the $3d$ electronic states of different symmetry of the Cu layer leads to the lifting of this degeneracy and to the opening of the band gap at the $\mathrm{K}$ point. Symmetry breaking is also reflected in the photoemission data in the region of the Cu $d$ bands. Along the $\Gamma-\mathrm{K}$ line, the group of the  $k$-vector in free-standing graphene is $C_{3v}$, such that carbon $p_z$-derived bands and those from Cu $d_{xz,yz}$ would not be allowed to interact. However, an avoided crossing between these states, indicative of symmetry reduction, is apparent in the data of (Fig.~\ref{ARPES}). The symmetry must therefore be further reduced, on account of the lattice mismatch leading to the moir\'e structure. 

This conclusion is supported by analyzing the weight of the contributions from different orbitals to the bands below and above the band gap as a function of energy and electron wave vector. The lower panel of Fig.~\ref{HYBRID} shows the weight of the $p_z^{C_{top}}$ + $d_{z^2}^{Cu(S)}$ (top) and the $p_z^{C_{fcc}}$ + $d_{xz,yz}^{Cu(S)}$ (bottom) hybridized states, represented by the magnitude of the symbols. It is clear that the $p_z$ + $d_{z^2}$ hybrid state dominate the bands below the band gap, while the $p_z$ + $d_{xz,yz}$ one dominates the bands above. This clearly shows how hybridization leads to sublattice breaking evidenced, for example, by the reduced symmetry of the LEED pattern in Figure~\ref{STM}. Comparing the graphene/Cu/Ir(111) system with the original graphene/Ir(111) interfaces, the interaction or hybridization between graphene valence band state and Ir $d$ orbitals is weaker compared to the Cu $d$ orbitals and hence the sublattice symmetry breaking is more pronounced in the latter case. This to our knowledge is the first case where such a detailed assignment of orbitals contributing to the states around the band gap has been achieved. Similar analysis for graphene on other substrates in which gap opening occurs will permit an understanding of the interplay of substrate and graphene states, and will shed light on the as yet elusive correlation between bonding strength, magnitude of band gap, and its consequences for the structural arrangement.

\textit{In summary}, STM and photoemission data from an intercalated layer in between graphene and Ir(111), interpreted on the basis of state-of-the-art DFT calculations, permit a detailed analysis of the changes in doping and band gap opening in the graphene $\pi$ bands upon intercalation, in terms of state-specific hybridization between the different Cu $d$ and graphene $\pi$ bands. The strong spatial and energy overlap of the valence band states leads to the lifting of the sublattice symmetry of the carbon atoms which manifests itself as an opening of the band gap at the Dirac point. Analyses of this kind permit to predict the arrangement of the electronic states of graphene in other graphene-metal interfaces.

\section*{Methods}

\textit{Preparation of graphene/Ir(111) and graphene/Cu/Ir(111).}
The graphene/Ir(111) system was prepared in ultrahigh vacuum system for ARPES experiments according to the recipe described in details in Refs.~\cite{NDiaye:2008qq,Coraux:2009,Voloshina:2013dq} via cracking of propylene: $T = 1100^\circ$\,C, $p = 5 \times 10^{-8}$\,mbar, $t = 30$\,min. This procedure leads to the single-domain graphene layer on Ir(111) of very high quality that was verified by means of LEED and STM (in the separate experiments). Intercalation of the 1\,ML-thick Cu layer and formation of the graphene/Cu/Ir(111) intercalation-like system was achieved via annealing of the graphene/Ir(111) sample with the thin pre-deposited copper layer on top. The process of intercalation was monitored by measuring of the ``live'' C\,$1s$ and Ir\,$4f$ spectra of the system and formation of the graphene/Cu/Ir(111) system was detected at $550^\circ$\,C. The quality and cleanness of the system in ARPES experiment was verified by LEED and XPS/ARPES, respectively. The base vacuum was better than $8 \times 10^{-11}$\,mbar during all experiments. All ARPES and STM measurements were performed at room temperature.

\textit{ARPES experiments.}
The ARPES measurements at photon energies of $h\nu=65$\,eV and $h\nu=94$\,eV were performed at the BESSY\,II storage ring (Berlin) in the photoemission station using PHOIBOS 100 2D-CCD hemispherical analyzer from SPECS. Experiments with He\,II radiation ($h\nu=40.81$\,eV) were performed in the SPECS demo lab using a FlexPS system with PHOIBOS 150 2D-CCD analyser. In both cases a 5-axis motorized manipulator was used, allowing for a precise alignment of the sample in $k$ space. The sample was azimuthally pre-aligned in such way that the tilt (BESSY\,II) or polar (FlexPS) scans were performed along the $\Gamma-\mathrm{K}$ direction of the graphene-derived BZ with the photoemission intensity on the channelplate images acquired along the direction perpendicular to $\Gamma-\mathrm{K}$. The final 3D data set of the photoemission intensity as a function of kinetic energy and two emission angles, $I(E_{kin},angle_1,angle_2)$, were then carefully analyzed.

\textit{STM experiments.}
The STM measurements were performed in constant current (CC) mode. In this case the topography of the sample, $z(x, y)$, is studied with the corresponding signal, tunneling current ($I_T$), used as an input for the feedback loop. The STM images were collected with SPM Aarhus 150 equipped with the KolibriSensor\texttrademark\ from SPECS~\cite{Torbruegge:2010cf,Voloshina:2013dq} with a Nanonis Control system. In all measurements, a sharp W-tip was used which was cleaned \textit{in situ} via Ar$^+$-sputtering. In the STM images the tunnelling bias voltage, $U_T$, is referenced to the sample and the tunnelling current, $I_T$, is collected by the tip, which is virtually grounded.

\textit{DFT calculations.}
The crystallographic model of graphene/Ir(111) and graphene/Cu/Ir(111) presented in Fig.~\ref{STM} was used in the DFT calculations, which were carried out using the projector augmented plane wave method~\cite{Blochl:1994}, a plane wave basis set with a maximum kinetic energy of $400$\,eV and the PBE exchange-correlation potential~\cite{Perdew:1996}, as implemented in the VASP program~\cite{Kresse:1994}. The long-range van der Waals interactions were accounted for by means of DFT-D2 approach proposed by Grimme~\cite{Grimme:2006}. The studied system is modelled using supercell, which has a $(9 \times 9)$ lateral periodicity and contains one layer of $(10 \times 10)$ graphene on a five-layer slab of metal atoms. Metallic slab replicas are separated by ca. $20$\,\AA\ in the surface normal direction. To avoid interactions between periodic images of the slab, a dipole correction is applied~\cite{Neugebauer:1992}. The surface Brillouin zone is sampled with a $7 \times 7 \times 1$ $k$-point mesh centered the $\Gamma$ point.

\bibliographystyle{nature}

\clearpage

\section*{Acknowledgements} 

This work has been supported by the European Science Foundation (ESF) under the EUROCORES Programme EuroGRAPHENE (Project ``SpinGraph''). E.\,N.\,V. and K.\,H. appreciate the support from the German Research Foundation (DFG) through the Priority Program (SPP) 1459 ``Graphene''. The computing facilities (ZEDAT) of the Freie Universit\"at Berlin and the High Performance Computing Network of Northern Germany (HLRN) are acknowledged for computer time. 

\section*{Author contribution}

H.V. and S.B. perform XPS/ARPES experiments at BESSY\,II. Y.S.D. perform STM and ARPES experiments in the laboratory. E.N.V. perform DFT calculations and STM simulations with help from R.E.O. H.V., S.B., K.H., E.N.V., R.E.O, T.K., A.T., Y.S.D. contribute in the treatment of the experimental and theoretical data and in interpretation of results. Y.S.D. wrote a manuscript with the contributions from all co-authors.

\section*{Additional information}

\subsection*{Competing financial interests}

The authors declare no competing financial interests.

\clearpage

\clearpage

\clearpage

\clearpage

\clearpage
\noindent
Supplementary material for manuscript:\\
\textbf{Understanding the origin of band gap formation in graphene on metals: graphene on Cu/Ir(111)}\\
\newline
H. Vita,$^1$ S. B\"ottcher,$^1$ K. Horn,$^1$ E. N. Voloshina,$^2$ R. E. Ovcharenko,$^2$ Th. Kampen,$^3$ A. Thissen,$^3$ and Yu. S. Dedkov$^3$\\
\newline
$^1$Fritz-Haber-Institut der Max-Planck-Gesellschaft, 14195 Berlin, Germany\\
$^2$Institut f\"ur Chemie, Humboldt Universit\"at zu Berlin, Brook-Taylor-Str. 2, 12489 Berlin, Germany\\
$^3$SPECS Surface Nano Analysis GmbH, Voltastra\ss e 5, 13355 Berlin, Germany
\newline
\newline
\textbf{List of tables:}
\\
\newline
\noindent\textbf{Table\,T1.} Distances (in \AA) between graphene and the underlying metal layer, Ir or Cu, for high-symmetry positions in graphene/Ir(111) and graphene/Cu/Ir(111) systems, respectively. See Fig.\,S1 for the explanation of the corresponding notations of the high-symmetry positions of the graphene/metal moir\'e structures.
\newline
\newline
\textbf{List of figures:}
\\
\newline
\noindent\textbf{Fig.\,S1.} Definitions of the high-symmetry positions of the graphene/Metal(111) moir\'e structures: The capital-letters marks are used for the definition of the high-symmetry places in the graphene/Metal(111) moir\'e structure where carbon atoms surround the corresponding adsorption places of the Metal(111) surface. In this case we have the following notations: (a) ATOP-position -- carbon atoms surround the metal atom of the top layer and are placed in the \textit{hcp} and \textit{fcc} hollow positions of the Metal(111) stack above (S-1) and (S-2) Metal-layers, respectively ($hcp-fcc$ position); (b) FCC-position -- carbon atoms surround the $fcc$ hollow site of the Metal(111) surface and are placed in the $top$ and $hcp$ hollow positions of the Metal(111) stack above (S) and (S-1) Metal-layers, respectively ($top-hcp$); (c) HCP-position -- carbon atoms surround the $hcp$ hollow site of the Metal(111) surface and are placed in the $top$ and $fcc$ hollow positions of the Metal(111) stack above (S) and (S-2) Metal-layers, respectively ($top-fcc$ position); and (d) BRIDGE-position -- carbon atoms are bridged by the Metal atom in the (S) layer.
\newline
\noindent\textbf{Fig.\,S2.} (a) and (b) show a series of the C\,$1s$ and Ir\,$4f$ photoelectron spectra fast-collected during annealing of the thin pre-deposited Cu layer on graphene/Ir(111) indicating the formation of the graphene/Cu/Ir(111) system. Temperature on the plot is growing from bottom to top and intercalation appears at $550^\circ$\,C. (c) and (d) show a fit of the C\,$1s$ and Ir\,$4f$ photoemission lines acquired before and after intercalation of Cu in graphene/Ir(111). Photon energy is $h\nu=400$\,eV.
\newline
\noindent\textbf{Fig.\,S3.} (a) and (b) show a series of the C\,$1s$ and Ir\,$4f$ photoelectron spectra extracted from Fig.\,S2 where the actual temperature is specified for every spectra.
\newline
\noindent\textbf{Fig.\,S4.} Carbon-atom projected total DOS for the graphene/Cu/Ir(111) system calculated by the tetrahedron method for the graphene supercell in the slab geometry. Inset shows the zoom of the DOS plot around band gap and $E_F$. The band gap region is marked by the shadow area.
\newline
\noindent\textbf{Fig.\,S5.} (a) Carbon-atom projected PDOSs for different high-symmetry places of the graphene/Cu/Ir(111) system. (b) Cu\,$3d$ orbital-projected DOSs for graphene/Cu/Ir(111).
\newline
\noindent\textbf{Fig.\,S6.} Simulated STM images of graphene/Cu/Ir(111) for different bias voltages. The energy range of integration corresponding to the respective bias voltage as well as the corrugations are marked on top of every image. The black rhombus marks the moir\'e unit cell and the corresponding symbols mark the high symmetry positions of the graphene/Cu/Ir(111) system.

\clearpage
\noindent\textbf{Table\,T1.} Distances (in \AA) between graphene and the underlying metal layer, Ir or Cu, for high-symmetry positions in graphene/Ir(111) and graphene/Cu/Ir(111) systems, respectively. See Fig.\,S1 for the explanation of the corresponding notations of the high-symmetry positions of the graphene/metal moir\'e structures.
\begin{table}
    \begin{tabular}{c|c|c}
    Position/System & gr/Ir(111) & gr/Cu/Ir(111) \\ \hline
    ATOP            & 3.581          & 3.122             \\
    FCC             & 3.280          & 2.893             \\
    HCP             & 3.274          & 3.006             \\
    BRIDGE          & 3.315          & 3.002             \\
    \end{tabular}
\end{table}

\clearpage
\begin{figure}
\includegraphics[scale=1.75]{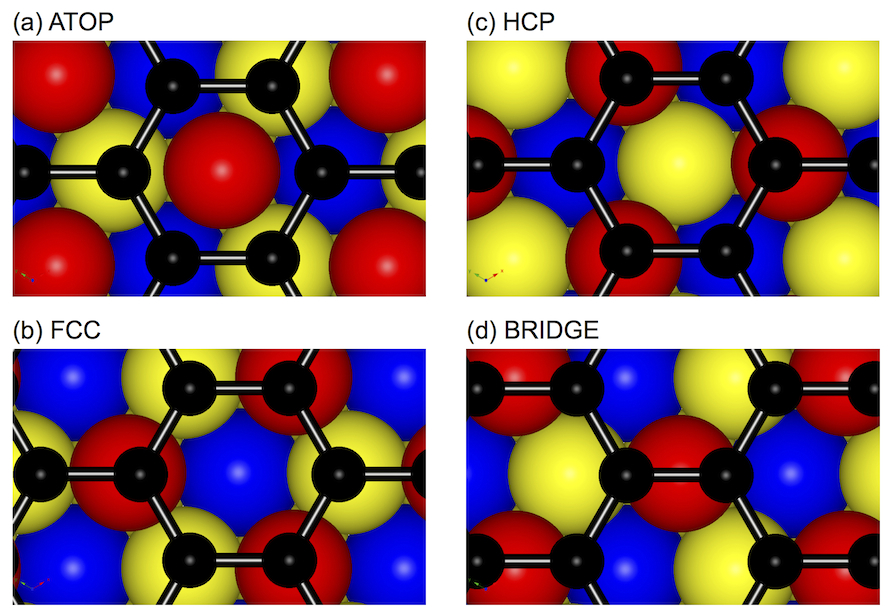}\\
\end{figure}
\noindent\textbf{Fig.\,S1.} (Definitions of the high-symmetry positions of the graphene/Metal(111) moir\'e structures: The capital-letters marks are used for the definition of the high-symmetry places in the graphene/Metal(111) moir\'e structure where carbon atoms surround the corresponding adsorption places of the Metal(111) surface. In this case we have the following notations: (a) ATOP-position -- carbon atoms surround the metal atom of the top layer and are placed in the \textit{hcp} and \textit{fcc} hollow positions of the Metal(111) stack above (S-1) and (S-2) Metal-layers, respectively ($hcp-fcc$ position); (b) FCC-position -- carbon atoms surround the $fcc$ hollow site of the Metal(111) surface and are placed in the $top$ and $hcp$ hollow positions of the Metal(111) stack above (S) and (S-1) Metal-layers, respectively ($top-hcp$); (c) HCP-position -- carbon atoms surround the $hcp$ hollow site of the Metal(111) surface and are placed in the $top$ and $fcc$ hollow positions of the Metal(111) stack above (S) and (S-2) Metal-layers, respectively ($top-fcc$ position); and (d) BRIDGE-position -- carbon atoms are bridged by the Metal atom in the (S) layer.

\clearpage
\begin{figure}
\includegraphics[scale=2.5]{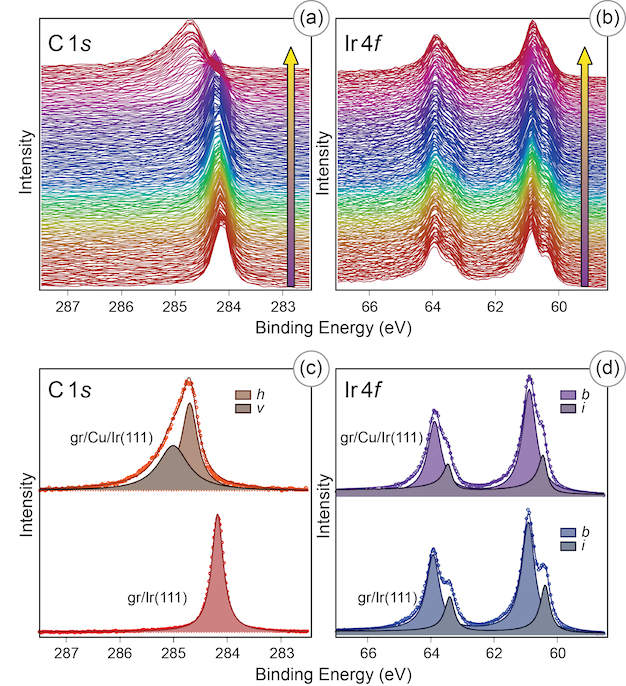}\\
\end{figure}
\noindent\textbf{Fig.\,S2.} (a) and (b) show a series of the C\,$1s$ and Ir\,$4f$ photoelectron spectra fast-collected during annealing of the thin pre-deposited Cu layer on graphene/Ir(111) indicating the formation of the graphene/Cu/Ir(111) system. Temperature on the plot is growing from bottom to top and intercalation appears at $550^\circ$\,C. (c) and (d) show a fit of the C\,$1s$ and Ir\,$4f$ photoemission lines acquired before and after intercalation of Cu in graphene/Ir(111). Photon energy is $h\nu=400$\,eV.

\clearpage
\begin{figure}
\includegraphics[scale=1.5]{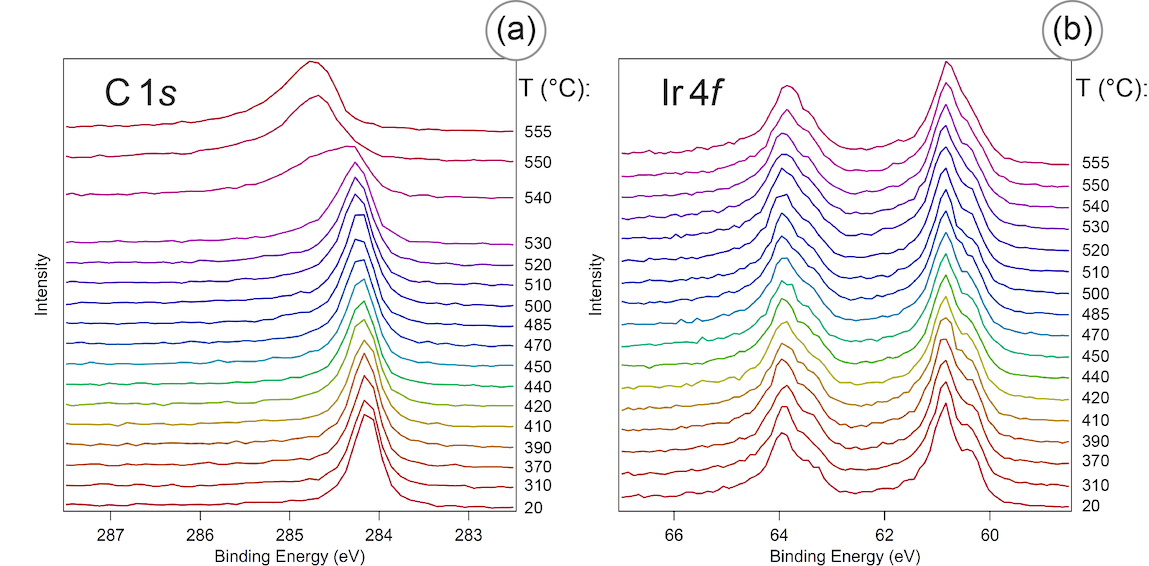}\\
\end{figure}
\noindent\textbf{Fig.\,S3.} (a) and (b) show a series of the C\,$1s$ and Ir\,$4f$ photoelectron spectra extracted from Fig.\,S2 where the actual temperature is specified for every spectra.

\clearpage
\begin{figure}
\includegraphics[scale=1.5]{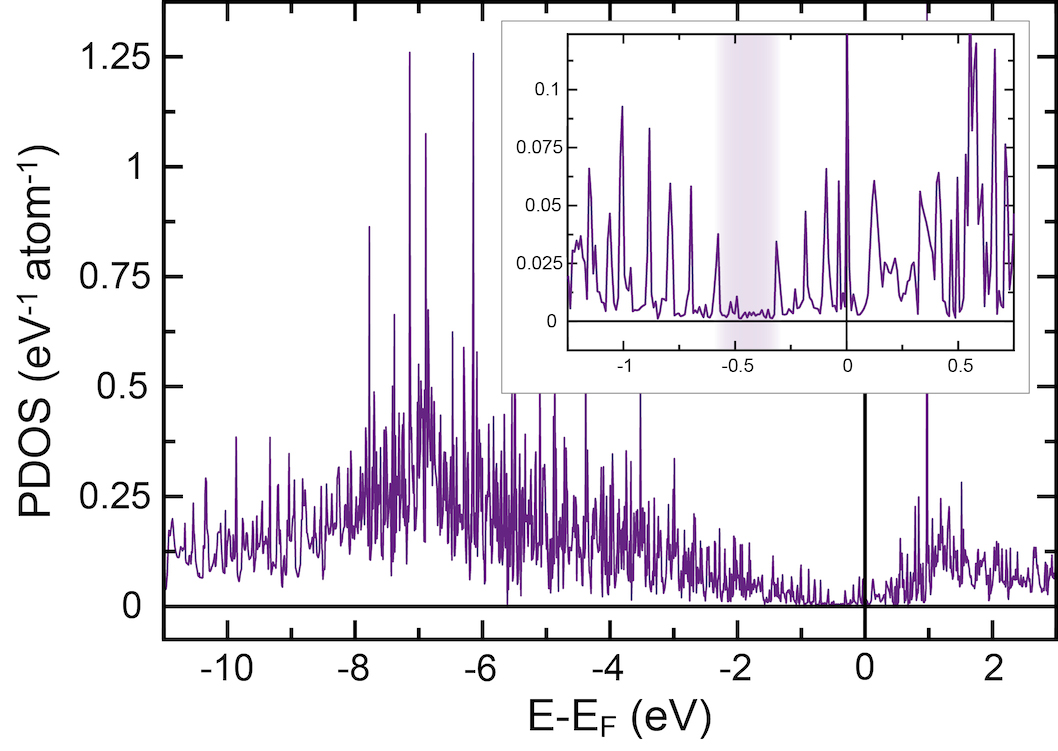}\\
\end{figure}
\noindent\textbf{Fig.\,S4.} Carbon-atom projected total DOS for the graphene/Cu/Ir(111) system calculated by the tetrahedron method for the graphene supercell in the slab geometry. Inset shows the zoom of the DOS plot around band gap and $E_F$. The band gap region is marked by the shadow area.

\clearpage
\begin{figure}
\includegraphics[scale=1.5]{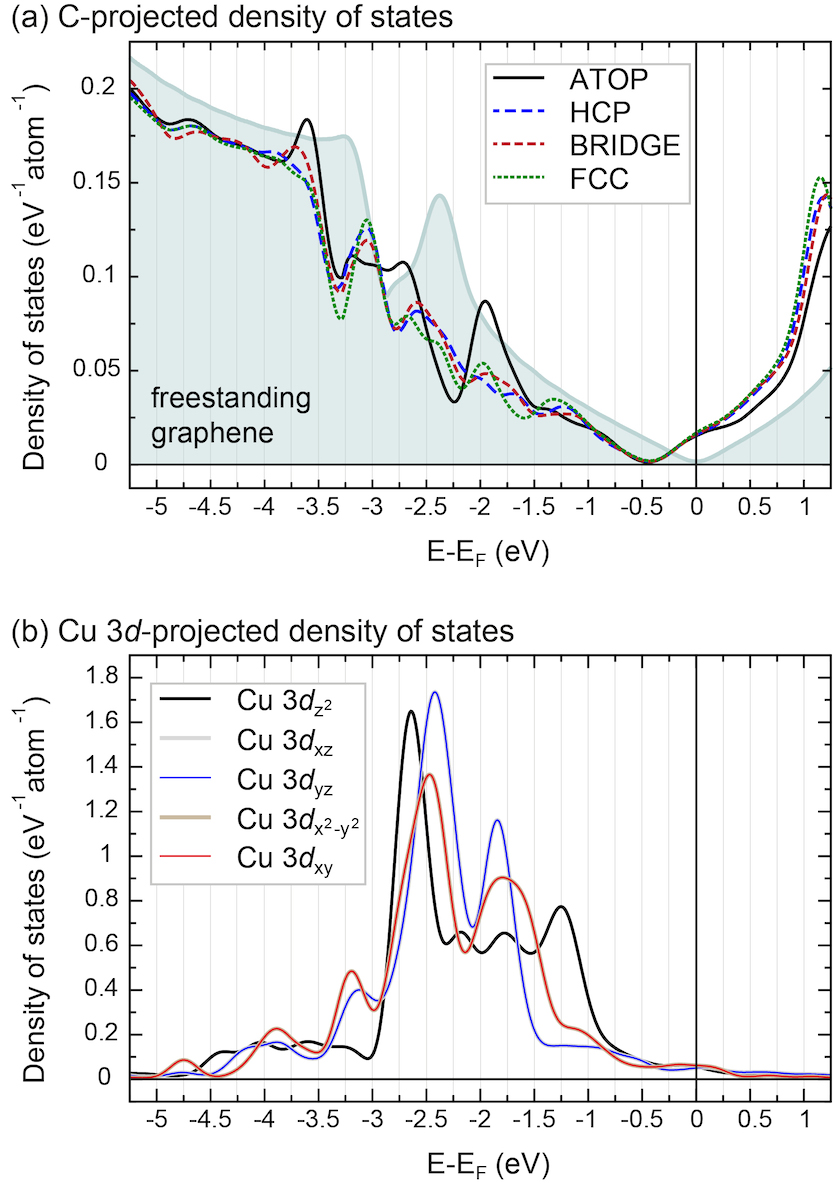}\\
\end{figure}
\noindent\textbf{Fig.\,S5.} (a) Carbon-atom projected PDOSs for different high-symmetry places of the graphene/Cu/Ir(111) system. (b) Cu\,$3d$ orbital-projected DOSs for graphene/Cu/Ir(111). 

\clearpage
\begin{figure}
\includegraphics[scale=1.5]{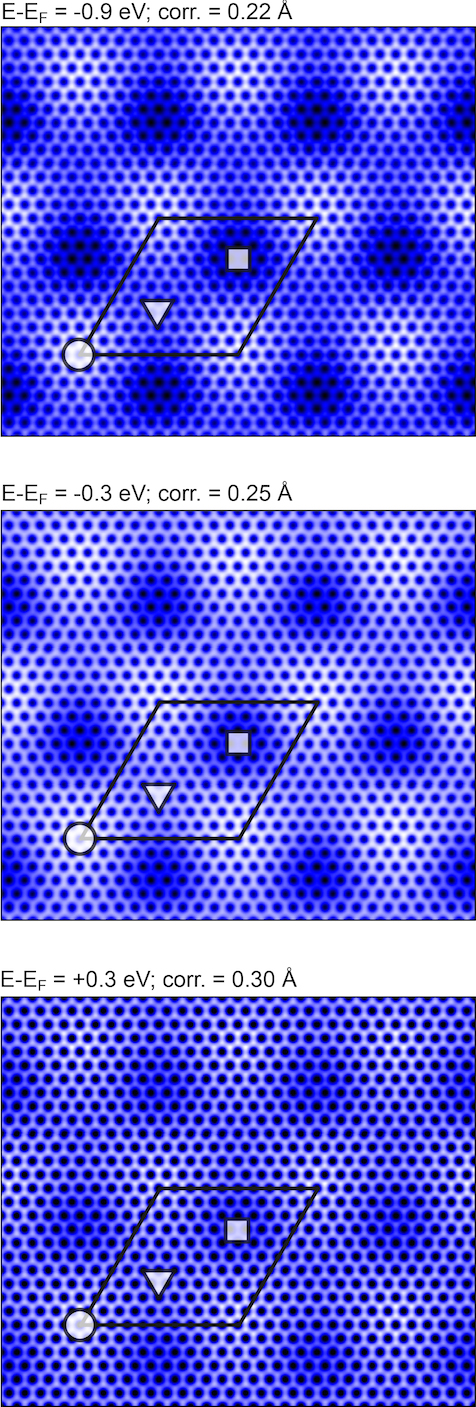}\\
\end{figure}
\noindent\textbf{Fig.\,S6.} Simulated STM images of graphene/Cu/Ir(111) for different bias voltages. The energy range of integration corresponding to the respective bias voltage as well as the corrugations are marked on top of every image. The black rhombus marks the moir\'e unit cell and the corresponding symbols mark the high symmetry positions of the graphene/Cu/Ir(111) system.

\end{document}